\documentclass[10pt,conference]{IEEEtran}
\IEEEoverridecommandlockouts
\usepackage[compress]{cite}
\usepackage[utf8]{inputenc}
\usepackage{amsmath}
\usepackage{amssymb}
\usepackage{mathrsfs}
\usepackage{enumerate}
\usepackage{adjustbox}
\usepackage{xcolor}
\usepackage{graphicx}
\usepackage{algorithm}
\usepackage{algpseudocode}
\usepackage{xpatch}
\usepackage{mathtools}
\usepackage{nicefrac}
\usepackage{float}
\floatstyle{plaintop}
\restylefloat{table}
\usepackage{cleveref}

\algdef{SE}[DOWHILE]{Do}{doWhile}{\algorithmicdo}[1]{\algorithmicwhile\ #1}%
\algrenewcommand\alglinenumber[1]{{\sffamily\footnotesize#1}}
\makeatletter
\xpatchcmd{\algorithmic}{\itemsep\z@}{\itemsep=.25ex plus2pt}{}{}
\makeatother

\usepackage{color}
\usepackage{soul}
\usepackage{environ}         
\usepackage{etoolbox}        
\usepackage[acronym]{glossaries}
\newacronym{1g}{1G}{first-generation}
\newacronym{4g}{4G}{fourth-generation}
\newacronym{5g}{5G}{fifth-generation}
\newacronym{6g}{6G}{sixth-generation}
\newacronym{mimo}{MIMO}{multiple-input multiple-output}
\newacronym{umi}{UMi}{Urban Micro-cellular}
\newacronym{ris}{RIS}{reconfigurable intelligent surface}
\newacronym{siso}{SISO}{single-input single-output}
\newacronym{mmimo}{mMIMO}{massive multiple-input multiple-output}
\newacronym{cfmmimo}{CF-mMIMO}{cell-free massive multiple-input-multiple-output}
\newacronym{isac}{ISAC}{integrated sensing and communication}
\newacronym{sumimo}{SU-MIMO}{single user MIMO}
\newacronym{mumimo}{MU-MIMO}{multi user MIMO}
\newacronym{embms}{eMBMS}{evolved Multimedia Broadcast and Multicast Service}
\newacronym{sca}{SCA}{successive convex approximation}
\newacronym{sinr}{SINR}{signal-to-interference-plus-noise ratio}
\newacronym{ula}{ULA}{uniform linear array}
\newacronym{uaf}{UatF}{\emph{use-and-then-forget}}
\newacronym{mcs}{MCS}{modulation and coding scheme}
\newacronym{dcc}{DCC}{dynamic cooperation clustering}
\newacronym{mrt}{MRT}{maximum ratio transmission}
\newacronym{ipmmse}{IP-MMSE}{improved partial MMSE}\newacronym{pmmse}{P-MMSE}{partial MMSE}
\newacronym{pzf}{P-ZF}{partial ZF}
\newacronym{zf}{ZF}{zero-forcing}
\newacronym{mr}{MR}{maximum ratio}
\newacronym{se}{SE}{spectral efficiency}
\newacronym{ase}{ASE}{aggregated spectral efficiency}
\newacronym{ee}{EE}{energy efficiency}
\newacronym{ap}{AP}{access point}
\newacronym{cpu}{CPU}{central processing unit}
\newacronym{uc}{UC}{user centric}
\newacronym{sse}{SumSE}{sum spectral efficiency}
\newacronym{mise}{MinSE}{minimum spectral efficiency}
\newacronym{asd}{ASD}{angular standard deviation}
\newacronym{adr}{ADR}{aggregated data rate}
\newacronym{embb}{eMBB}{enhanced mobile broadband}
\newacronym{mmtc}{mMTC}{massive machine type communications}
\newacronym{urllc}{URLLC}{ultra reliable low latency communications}
\newacronym{csi}{CSI}{channel state information}
\newacronym{pmi}{PMI}{precoding matrix indicator}
\newacronym{ri}{RI}{rank indicator}
\newacronym{csi-rs}{CSI-RS}{CSI-reference signal}
\newacronym{qos}{QoS}{quality of service}
\newacronym{cri}{CRI}{CSI-RS resource indicator}
\newacronym{bs}{BS}{base station}
\newacronym{re}{RE}{resource element}
\newacronym{mmwave}{mmWave}{millimeter-wave}
\newacronym{umwave}{$\mu$mWaves}{micrometer waves}
\newacronym{rnn}{RNN}{recurrent neural network}
\newacronym{cnn}{CNN}{convolutional neural network}
\newacronym{ngmn}{NGMN}{next-generation mobile network}
\newacronym{lte}{LTE}{Long Term Evolution}
\newacronym{lte-a}{LTE-A}{Long Term Evolution Advanced}
\newacronym{5gnr}{5G NR}{5G New Radio}
\newacronym{mm}{MM}{mixed mode}
\newacronym{cdf}{CDF}{cumulative distribution function}
\newacronym{phy}{PHY}{physical}
\newacronym{mac}{MAC}{medium access control}
\newacronym{3gpp}{3GPP}{3rd Generation Partnership Project}
\newacronym{fdd}{FDD}{frequency division duplexing}
\newacronym{tdd}{TDD}{time division duplexing}
\newacronym{ofdm}{OFDM}{orthogonal frequency division multiplexing}
\newacronym{ss}{SS}{synchronization signal} 
\newacronym{pss}{PSS}{primary synchronization signal} 
\newacronym{sss}{SSS}{secondary synchronization signal} 
\newacronym{pbch}{PBCH}{physical broadcast channel} 
\newacronym{dmrs}{DMRS}{demodulation reference signal} 
\newacronym{gnb}{gNB}{next generation nodeB} 
\newacronym{rsrp}{RSRP}{reference signal received power} 
\newacronym{rrm}{RRM}{radio resource management} 
\newacronym{srs}{SRS}{sounding reference signal} 
\newacronym{ran}{RAN}{radio access network} 
\newacronym{nn}{NN}{neural network} 
\newacronym{ue}{UE}{user equipment} 
\newacronym{awgn}{AWGN}{additive white Gaussian noise} 
\newacronym{epa}{EPA}{Extended Pedestrian A model}
\newacronym{eva}{EVA}{Extended Vehicular A model}
\newacronym{etu}{ETU}{Extended Typical Urban model}
\newacronym{tdl}{TDL}{tapped delay line}
\newacronym{cdl}{CDL}{clustered delay line}
\newacronym{uma}{UMa}{urban macro-cell}
\newacronym{isd}{ISD}{inter-site distance}
\newacronym{nlos}{NLOS}{non-line of sight}
\newacronym{los}{LOS}{line of sight}
\newacronym{o2o}{O2O}{outdoor-to-outdoor}
\newacronym{o2i}{O2I}{outdoor-to-indoor}
\newacronym{ul}{UL}{uplink}
\newacronym{dl}{DL}{downlink}
\newacronym{ls}{LS}{least squares}
\newacronym{mmse}{MMSE}{minimum mean square error}
\newacronym{snr}{SNR}{signal-to-noise ratio}
\newacronym{mse}{MSE}{mean square error}
\newacronym{nr}{NR}{New Radio}
\newacronym{prb}{PRB}{physical resource block}
\newacronym{scs}{SCS}{subcarrier spacing}
\newacronym{bler}{BLER}{block error rate}
\newacronym{smmmra}{SMMMRA}{subgroup multicast \gls{mamimo} resource allocation}
\newacronym{mmf}{MMF}{max-min fairness}
\newacronym{smmu}{SMMU}{subgroups of multicast \gls{mamimo} users}
\newacronym{gsmma}{GSMMA}{greedy subgroup multicast \gls{mamimo} algorithm}
\newacronym{apa}{APA}{adaptive power allocation}
\newacronym{ms}{MS}{mobile station}
\newacronym{cb}{CB}{conjugate beamforming}
\newacronym{ncb}{NCB}{normalized conjugate beamforming}
\newacronym{ecb}{ECB}{enhanced conjugate beamforming}
\usepackage{flushend}

\usepackage{tikz}
\usetikzlibrary{calc,shapes.geometric}
\usepackage{changepage}
\usetikzlibrary{fadings}
\tikzfading[name=fade out, 
    inner color=transparent!0,
    outer color=transparent!100]
\usetikzlibrary{patterns}
\usetikzlibrary{shadows.blur}
\usetikzlibrary{shapes}

\ifCLASSOPTIONcompsoc
    \usepackage[caption=false, font=normalsize, labelfont=sf, textfont=sf, subrefformat=parens, labelformat=parens]{subfig}
\else
\usepackage[caption=false, font=footnotesize, subrefformat=parens, labelformat=parens]{subfig}
\fi

\makeatletter
\newcommand\fs@betterruled{%
  \def\@fs@cfont{\bfseries}\let\@fs@capt\floatc@ruled
  \def\@fs@pre{\vspace*{5pt}\hrule height.8pt depth0pt \kern2pt}%
  \def\@fs@post{\kern2pt\hrule\relax}%
  \def\@fs@mid{\kern2pt\hrule\kern2pt}%
  \let\@fs@iftopcapt\iftrue}
\floatstyle{betterruled}
\restylefloat{algorithm}
\makeatother

\begin{document}
\title{Conjugate Beamforming Variants for Multicasting in Cell-Free Massive MIMO Systems 
\thanks{This work has been supported by grants SOFIA-WIND (PID2023-147305OB-C33, funded by MICIU/AEI/10.13039/501100011033 and ERDF, EU), brAIn5G (PID2024-161515OA-I00, funded by MICIU/AEI/10.13039/501100011033 and FEDER), POLIGRAPH (PID2022-136887NB-I00, funded by MCIU/AEI/10.13039/501100011033), and the project AI-XCAST6G (ref. F1263, funded by the IMPULSO program for research at the Rey Juan Carlos University).
}}

\author{\IEEEauthorblockN{Alejandro de la Fuente\IEEEauthorrefmark{1}, Adrián Espinosa\IEEEauthorrefmark{1}, Jan García-Morales\IEEEauthorrefmark{1}, Guillem Femenias\IEEEauthorrefmark{2}, Felip Riera-Palou\IEEEauthorrefmark{2}}
\IEEEauthorblockA{\IEEEauthorrefmark{1}Dept. of Signal Theory and Communications, University Rey Juan Carlos, 28942 Fuenlabrada (Madrid), Spain}
\IEEEauthorblockA{\IEEEauthorrefmark{2}SECOM Group and Artificial Intelligence Research Institute (IAIB), University of the Balearic Islands (UIB), Spain}
Email: alejandro.fuente@urjc.es}

\maketitle

\begin{abstract}
This paper studies scalable conjugate beamforming (CB) variants for physical-layer multicasting in cell-free massive multiple-input multiple-output (CF-mMIMO) systems. Focusing on fully distributed precoding, we analyze classical CB, normalized CB (NCB), and enhanced CB (ECB) within a subgroup-centric multicast framework. Multicast users are
partitioned into subgroups based on large-scale fading similarity, which enables composite channel estimation, pilot reuse, and distributed precoding with low complexity. The performance of the different CB variants is evaluated in terms of aggregated spectral efficiency (ASE) under representative user geometries, including uniformly distributed users, spatially clustered deployments, and heterogeneous scenarios combining hotspots with more dispersed users. Monte Carlo simulations reveal a strong spatial geometry-dependent behavior: unicast transmission is preferable in uniform deployments, while subgroup-based multicasting becomes essential in clustered and heterogeneous scenarios. Among the CB-based precoders, NCB offers a robust performance-complexity trade-off across most scenarios, whereas ECB provides additional gains only when sufficient channel hardening is present. These results provide practical insights into the selection of low-complexity distributed precoders and multicast transmission modes in CF-mMIMO systems supporting broadband and multimedia services.
\end{abstract}

\glsresetall
\section{Introduction}

Cell-free massive multiple-input multiple-output (CF-mMIMO) has emerged as one of the most promising physical-layer technologies for beyond-5G and 6G systems thanks to its ability to deliver uniformly high spectral efficiency (SE), macro-diversity, and improved reliability across large coverage areas \cite{2016Marzetta,2024Ngo,2021Demir}. In CF-mMIMO systems, a large number of distributed access points (APs) jointly serve every user equipment (UE) using coherent transmission under time-division duplexing (TDD), eliminating cell boundaries and mitigating inter-cell interference. This architecture is particularly suitable for dense heterogeneous deployments, where user distributions exhibit a wide range of spatial geometries.

In parallel, physical-layer multicasting has gained considerable attention due to its relevance for multimedia delivery, public safety, and emerging 6G use cases in which common multicast data streams are transmitted to multiple UEs 
\cite{2025Ericsson,2016delaFuente,2017Araniti}. Classical approaches for delivering the same information to multiple UEs include: (i) a single multicast transmission serving all users, and (ii) individual unicast transmissions to each UE.
In CF-mMIMO, however, these approaches become inefficient: a single multicast stream is bottlenecked by the worst UE in the entire network, whereas unicast suffers from severe pilot contamination and excessive resource consumption when the number of users grows large.

To overcome these limitations, recent work has proposed subgroup‑centric multicast transmission, where users are partitioned into multicast subgroups based on similarities in their large-scale channel statistics \cite{2022delaFuente,2024delaFuente}. Users within the same subgroup share an uplink (UL) pilot sequence and a downlink (DL) precoder,
enabling more accurate channel estimation, reduced pilot contamination, and enhanced SE. This design leverages the fact that UEs experiencing similar propagation conditions benefit from a common DL transmission, as observed in both massive MIMO \cite{2018Sadeghi} and CF-mMIMO \cite{2017Doan}.

Scalability in CF-mMIMO systems is not inherently tied to a specific precoding architecture, as both distributed and centralized schemes can be designed to scale efficiently. Rather, it depends on factors such as signaling overhead, computational complexity, and fronthaul requirements. In this context, conjugate beamforming (CB), implemented locally at each AP using only its own channel estimates, entails minimal coordination and signaling, making it particularly attractive from a scalability perspective \cite{2021Interdonato}.
Several distributed CB variants have been proposed to improve channel hardening and mitigate the variability of the effective DL channel gain. These include normalized CB (NCB), which scales the signal by the inverse estimated channel norm to reduce gain fluctuations, and enhanced CB (ECB), which applies a squared-norm normalization for further stabilization of the effective channel gain \cite{2020Femenias,2021Interdonato}.
While their performance has been analyzed in unicast settings, their behavior in multicast CF-mMIMO systems, especially under different spatial user distributions, remains poorly understood.

Recent studies demonstrate that the spatial distribution of users plays a fundamental role in multicast performance \cite{2024delaFuente}. Uniform UE deployments often favor unicast transmissions, whereas clustered geometries highlight the benefits of multicast subgrouping due to severe intra‑cluster pilot contamination. Moreover, heterogeneous deployments have also been analyzed, where some UEs are uniformly distributed over the coverage area while others are concentrated in hotspots, which is highly relevant for realistic CF-mMIMO scenarios. 

\vspace{1mm}
\textbf{Contributions:}  
Building upon these foundations, this paper presents a comprehensive evaluation of three scalable distributed precoders, CB, NCB, and ECB, within a subgroup-centric CF-mMIMO multicasting framework. The key contributions are:
\begin{itemize}
    \item We develop a unified subgroup-centric CF-mMIMO simulation framework combining composite channel estimation, distributed precoding, and adaptive AP cooperation clustering.

    \item We evaluate the performance of CB, NCB, and ECB under three representative user spatial geometries: uniform, spatially clustered, and heterogeneous deployments with mixed hotspot–uniform regions.

    \item We quantify the performance gains and trade-offs of instantaneous normalization (NCB) and squared-norm normalization (ECB) relative to classical CB, providing insights into the regimes where each variant is preferable. 

    \item We present a spatial geometry-dependent analysis of the optimal multicast mode (unicast, single multicast, or subgroup multicast) and outline practical guidelines for low-complexity distributed multicasting in CF-mMIMO networks.
\end{itemize}

Overall, this work extends prior studies on CF-mMIMO multicasting \cite{2024delaFuente,2017Doan} by conducting the first in-depth comparison of several CB-based distributed precoders in realistic, heterogeneous multicast deployments. 

\section{System Model}

We consider a CF-mMIMO network operating in TDD and composed of a central processing unit
(CPU) connected via ideal fronthaul links to $L$ distributed APs. Each AP is equipped with $N$ antennas and jointly serves $K$ single-antenna UEs over the same time-frequency
resources. The set of UEs is denoted by $\mathcal{K}=\{1,\dots,K\}$ and the set of APs by $\mathcal{L}=\{1,\dots,L\}$. Each coherence block contains $\tau_{\mathrm{c}}$ samples, divided into $\tau_{\mathrm{p}}$ UL pilot symbols and $\tau_{\mathrm{d}}=\tau_{\mathrm{c}}-\tau_{\mathrm{p}}$ DL data symbols.

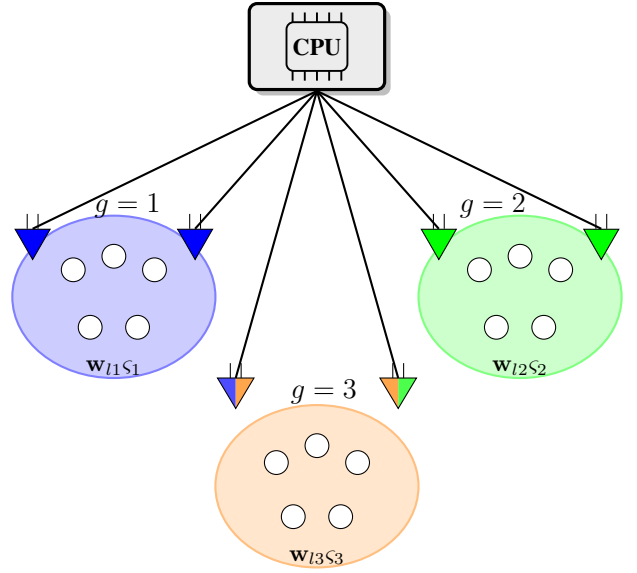
\begin{figure}[t]
\centering
\begin{tikzpicture}[scale=0.9,transform shape,
    user/.style={circle,draw,inner sep=1.1pt,minimum size=10pt,fill=white},
    ap/.style={regular polygon,regular polygon sides=3,draw,fill=#1,
               inner sep=3pt,minimum size=13pt,shape border rotate=180},       
    cpu/.style={rectangle,draw=black,very thick,fill=gray!10,
                rounded corners=2pt,minimum width=2.6cm,
                minimum height=1.1cm},
    arrow/.style={-stealth,thick},
    groupcircle/.style={ellipse,draw=#1!50,fill=#1!20,thick,
                        minimum width=3.0cm,minimum height=2.4cm},
    link/.style={thick}
    ]

\node[rectangle,draw=black,very thick,
      fill=gray!15,minimum width=2cm,minimum height=1.1cm,
      rounded corners=3pt,
      drop shadow]
      (CPU) at (0,3.7)
      {
        \begin{tikzpicture}[scale=0.18]
            \draw[thick] (0,0) rectangle (5,4);
            \foreach \x in {0.5,1.5,2.5,3.5,4.5}{
                \draw[thick] (\x,4) -- (\x,4.8);
                \draw[thick] (\x,0) -- (\x,-0.8);
            }
        \end{tikzpicture}      
      };
\node at (0,3.7)       {\textbf{CPU}};
\node[groupcircle=blue] (SG1) at (-3,0) {};
\node at (-2.8,1.35) {\large $g = 1$};

\node[user] at ($(SG1.center)+(-0.6,0.4)$) {};
\node[user] at ($(SG1.center)+(0.0,0.6)$) {};
\node[user] at ($(SG1.center)+(0.6,0.4)$) {};
\node[user] at ($(SG1.center)+(-0.35,-0.45)$) {};
\node[user] at ($(SG1.center)+(0.35,-0.45)$) {};

\node at ($(SG1.center)+(0,-1.05)$)
{$\mathbf{w}_{l1}\varsigma_{1}$};

\node[groupcircle=green] (SG2) at (3,0) {};
\node at (2.6,1.35) {\large $g = 2$};

\node[user] at ($(SG2.center)+(-0.6,0.4)$) {};
\node[user] at ($(SG2.center)+(0.0,0.6)$) {};
\node[user] at ($(SG2.center)+(0.6,0.4)$) {};
\node[user] at ($(SG2.center)+(-0.35,-0.45)$) {};
\node[user] at ($(SG2.center)+(0.35,-0.45)$) {};

\node at ($(SG2.center)+(0,-1.05)$)
{$\mathbf{w}_{l2}\varsigma_{2}$};

\node[groupcircle=orange] (SG3) at (0,-2.8) {};
\node at (0.1,-1.4) {\large $g = 3$};

\node[user] at ($(SG3.center)+(-0.6,0.35)$) {};
\node[user] at ($(SG3.center)+(0.0,0.6)$) {};
\node[user] at ($(SG3.center)+(0.6,0.35)$) {};
\node[user] at ($(SG3.center)+(-0.35,-0.45)$) {};
\node[user] at ($(SG3.center)+(0.35,-0.45)$) {};

\node at ($(SG3.center)+(0,-1.05)$)
{$\mathbf{w}_{l3}\varsigma_{3}$};


\node[ap=blue] (AP11) at (-4.2,0.85) {};
\node[ap=blue] (AP12) at (-1.8,0.85) {};

\draw ($(AP11.north)+(0.08,0)$) -- ++(0,0.22);
\draw ($(AP11.north)+(-0.08,0)$) -- ++(0,0.22);
\draw ($(AP12.north)+(0.08,0)$) -- ++(0,0.22);
\draw ($(AP12.north)+(-0.08,0)$) -- ++(0,0.22);

\node[ap=green] (AP21) at (1.8,0.85) {};
\node[ap=green] (AP22) at (4.2,0.85) {};

\draw ($(AP21.north)+(0.08,0)$) -- ++(0,0.22);
\draw ($(AP21.north)+(-0.08,0)$) -- ++(0,0.22);
\draw ($(AP22.north)+(0.08,0)$) -- ++(0,0.22);
\draw ($(AP22.north)+(-0.08,0)$) -- ++(0,0.22);

\coordinate (AP31low)  at (-1.2,-1.65);   
\coordinate (AP31L)    at (-1.47,-1.2);   
\coordinate (AP31R)    at (-0.93,-1.2);   
\coordinate (AP31mid)  at (-1.2,-1.2);   

\fill[blue!70]
  (AP31low) -- (AP31L) -- (AP31mid) -- cycle;

\fill[orange!70]
  (AP31low) -- (AP31mid) -- (AP31R) -- cycle;

\draw[thin]
  (AP31low) -- (AP31L) -- (AP31R) -- cycle;

\coordinate (AP32low)  at (1.2,-1.65);   
\coordinate (AP32L)    at (1.47,-1.2);   
\coordinate (AP32R)    at (0.93,-1.2);   
\coordinate (AP32mid)  at (1.2,-1.2);   

\fill[green!70]
  (AP32low) -- (AP32L) -- (AP32mid) -- cycle;

\fill[orange!70]
  (AP32low) -- (AP32mid) -- (AP32R) -- cycle;

\draw (AP32low) -- (AP32L) -- (AP32R) -- cycle;  

\draw ($(AP31mid.north)+(0.08,0)$) -- ++(0,0.22);
\draw ($(AP31mid.north)+(-0.08,0)$) -- ++(0,0.22);
\draw ($(AP32mid.north)+(0.08,0)$) -- ++(0,0.22);
\draw ($(AP32mid.north)+(-0.08,0)$) -- ++(0,0.22);

\draw[link] (CPU.south) -- (AP11.north);
\draw[link] (CPU.south) -- (AP12.north);
\draw[link] (CPU.south) -- (AP21.north);
\draw[link] (CPU.south) -- (AP22.north);
\draw[link] (CPU.south) -- (AP31mid.north);
\draw[link] (CPU.south) -- (AP32mid.north);

\end{tikzpicture}

\caption{Subgroup-centric CF-mMIMO multicast architecture. Users are partitioned into multicast subgroups according to large-scale fading similarity. Each subgroup is served by a cooperative cluster of distributed APs using a single multicast precoder $\mathbf{w}_{lg}\varsigma_{g}$. Inverted triangles denote the APs, while circles represent the UEs. Bicolor APs indicate dynamic cooperation, where an AP may serve multiple multicast subgroups. At the same time, a central CPU coordinates all APs through ideal fronthaul links.}
\label{fig:systemmodel}
\end{figure}

\subsection{Channel Model}

A spatially correlated Rayleigh block-fading channel is assumed. The channel between AP $l$ and UE $k$ is characterized as
\begin{equation}
    \mathbf{h}_{lk} \sim \mathcal{CN}(\mathbf{0}, \mathbf{R}_{lk}),
\end{equation}
where $\mathbf{R}_{lk}\in\mathbb{C}^{N\times N}$ is the positive semi-definite spatial covariance matrix. The average large-scale gain is
$\beta_{lk}=\mathrm{tr}(\mathbf{R}_{lk})/N$. It can be safely assumed that the channels of different AP-UE pairs are independent:
\begin{equation}
    \mathbb{E}\{\mathbf{h}_{l'k'} \mathbf{h}_{lk}^{H}\}= \mathbf{0},
    \quad \forall (l',k')\neq (l,k).
\end{equation}

Let us denote the global channel of UE $k$ as
\begin{equation}
    \mathbf{h}_k = 
    \begin{bmatrix}
        \mathbf{h}_{1k}^{\mathrm{T}} &
        \dots &
        \mathbf{h}_{Lk}^{\mathrm{T}}
    \end{bmatrix}^{\mathrm{T}}
    \sim \mathcal{CN}(\mathbf{0},\mathbf{R}_k),
\end{equation}
with $\mathbf{R}_k=\mathrm{blkdiag}(\mathbf{R}_{1k},\dots,\mathbf{R}_{Lk})$.
Covariance matrices vary on a large-scale fading timescale and are
assumed known at APs and CPU.

\subsection{Multicast Subgrouping}

The $K$ UEs requesting the same multicast service are partitioned into $G$ disjoint multicast subgroups $\mathcal{G}=\{1,\dots,G\}$, where $\mathcal{K}_g$ denotes the set of UEs in subgroup $g$. The UE partitioning relies on the similarity of the large-scale fading vectors
\begin{equation}
    \boldsymbol{\beta}_{k} = 
    [\beta_{1k},\dots,\beta_{Lk}]^{\mathrm{T}},
\end{equation}
which reflect the spatial geometry of each UE with respect to the APs. A K-means clustering rule groups UEs with similar $\boldsymbol{\beta}_k$ into the same subgroup \cite{2018Riera}, ensuring they experience similar propagation conditions and can share the same UL pilot and DL precoder.

Each subgroup $g$ is served by a subset of APs $\mathcal{L}_g\subseteq\mathcal{L}$ according to a subgroup-centric dynamic cooperation clustering (DCC) strategy. For each pilot index, each AP selects the subgroup for which it observes the strongest common
average channel gain $K_g^{-1}\!\sum_{k\in\mathcal{K}_g}\beta_{lk}$ \cite{2024delaFuente}.

\subsection{Uplink Channel Estimation}

All UEs in subgroup $g$ transmit the same pilot sequence
$\boldsymbol{\psi}_g\in\mathbb{C}^{\tau_{\mathrm{p}}}$, with
$\|\boldsymbol{\psi}_g\|^2=1$. Due to pilot reuse across subgroups,
pilot contamination may occur when $\boldsymbol{\psi}_c = \boldsymbol{\psi}_g$ for $c\neq g$.

The $N\times \tau_{\mathrm{p}}$ UL pilot signal received at AP $l$ is
\begin{equation}
    \mathbf{Y}_l
    = \sqrt{\tau_{\mathrm{p}} P_{\mathrm{p}}}
    \sum_{g=1}^G \sum_{k\in\mathcal{K}_g}
    \mathbf{h}_{lk}\boldsymbol{\psi}_g^{\mathrm{T}}
    + \mathbf{N}_l,
\end{equation}
where $P_{\mathrm{p}}$ is the pilot power and $\mathbf{N}_l$ is \gls{awgn} whose entries are i.i.d. and distributed as $\mathcal{CN}(0,\sigma_u^2)$.

Projecting $\mathbf{Y}_l$ onto $\boldsymbol{\psi}_g^{\!*}$ yields
\begin{equation}
    \mathbf{y}_{l}^g =
    \sqrt{\tau_{\mathrm{p}} P_{\mathrm{p}}}\sum_{k\in\mathcal{K}_g}\mathbf{h}_{lk}
    + \sqrt{\tau_{\mathrm{p}} P_{\mathrm{p}}}
    \sum_{\substack{c\neq g\\\boldsymbol{\psi}_c=\boldsymbol{\psi}_g}}
    \sum_{i\in\mathcal{K}_c}\mathbf{h}_{li}
    + \mathbf{n}_{lg}.
\end{equation}

APs cannot separate co-pilot UEs, resulting in the so-called pilot contamination effect.Thus, they estimate the \emph{composite channel} as
\begin{equation}
    \mathbf{h}_{l}^g \triangleq
    \frac{\sqrt{\tau_{\mathrm{p}} P_{\mathrm{p}}}}{K_g}\sum_{k\in\mathcal{K}_g}\mathbf{h}_{lk}
    \sim \mathcal{CN}(\mathbf{0},\mathbf{R}^g_l),
\end{equation}
where
\begin{equation}
    \mathbf{R}^g_l = \frac{\tau_{\mathrm{p}} P_{\mathrm{p}}}{K_g^2}
    \sum_{k\in\mathcal{K}_g}
    \mathbf{R}_{lk}.
\end{equation}

The \gls{mmse} estimate at AP $l$ is
\begin{equation}
    \hat{\mathbf{h}}_{l}^g
    = K_g\, \mathbf{R}^g_l
      \boldsymbol{\Gamma}_g^{-1}\mathbf{y}_{l}^g,
\end{equation}
with
\begin{equation}
    \boldsymbol{\Gamma}_g =
    \tau_{\mathrm{p}} P_{\mathrm{p}}
    \sum_{\substack{c:\,\boldsymbol{\psi}_c=\boldsymbol{\psi}_g}}
    \sum_{i\in\mathcal{K}_c}\mathbf{R}_{li}
    + \sigma_u^2\mathbf{I}_N.
\end{equation}

\subsection{Downlink Data Transmission and Spectral Efficiency}

The CPU maps the multicast stream into $G$ independent symbols
$\{\varsigma_g\}$, one per subgroup. AP $l$ transmits
\begin{equation}
    \mathbf{x}_{lg} = \mathbf{D}_{lg}\mathbf{w}_{lg}\varsigma_g,
\end{equation}
where $\mathbf{w}_{lg}$ is the DL precoder designed
from the local estimate $\hat{\mathbf{h}}_{l}^g$, and the set of diagonal matrices $\mathbf{D}_{lg} \in \mathbb{C}^{N\times N}$ is used to describe which APs communicate with which multicast subgroups, and are given by $\mathbf{D}_{lg} \!=\! \mathbf{I}_N$ if $l \in \mathcal{L}_{g}$, or $\mathbf{D}_{lg} \!=\!\boldsymbol{0}_{N\times N}$, otherwise.

The received signal at UE $k\in\mathcal{K}_g$ is
\begin{equation}
    y_k =
    \sum_{l\in\mathcal{L}}
    \mathbf{h}_{lk}^{H}\mathbf{D}_{lg}\mathbf{w}_{lg}\varsigma_g
    + \sum_{c\neq g}
    \sum_{l\in\mathcal{L}}
    \mathbf{h}_{lk}^{H}\mathbf{D}_{lc}\mathbf{w}_{lc}\varsigma_c
    + n_k.
\end{equation}

Assuming that UEs have access to the average value of the effective precoded channel, a deterministic quantity that can be easily obtained in practice \cite{2021Demir}, the channel hardening bounding technique can be applied to obtain the achievable SE of UE $k$ as 
\begin{equation}
    \mathrm{SE}_k =
    \Big(1 - \frac{\tau_{\mathrm{p}}}{\tau_{\mathrm{c}}}\Big)
    \log_2\!\left(1+\gamma_k\right),
\end{equation}
where the effective \gls{sinr} is
\begin{equation}
\resizebox{\columnwidth}{!}{$
\gamma_k =
\frac{\Big|\sum_{l}\mathbb{E}\{\mathbf{h}_{lk}^{H}\mathbf{D}_{lg}\mathbf{w}_{lg}\}\Big|^2}
{\sum_{c=1}^{G}\mathbb{E}\!\left[\Big|
\sum_{l}\mathbf{h}_{lk}^{H}\mathbf{D}_{lc}\mathbf{w}_{lc}
\Big|^2\right] -
\Big|\sum_{l}\mathbb{E}\{\mathbf{h}_{lk}^{H}\mathbf{D}_{lg}\mathbf{w}_{lg}\}\Big|^2
+ \sigma_d^2 }.$}
\end{equation}

Since all users in subgroup $g$ must be able to decode the same symbol, the subgroup SE is
\begin{equation}
    \mathrm{SE}_g = \min_{k\in\mathcal{K}_g}\mathrm{SE}_k.
\end{equation}

\section{Distributed Conjugate Beamforming Variants}

In this section, we present three fully distributed precoding schemes based on CB. All APs construct their DL precoders using only the local composite channel estimate $\hat{\mathbf{h}}_{l}^g$ associated with multicast subgroup $g$, without requiring CSI exchange across the network. This decentralization ensures excellent scalability and enables simple per‑AP power control. The three CB-type schemes explored in this work are: (i) classical CB, (ii) NCB, and (iii) ECB. All techniques are compatible with the subgroup-centric design described in Sec.~II.

Let $\mathbf{v}_{lg}\in\mathbb{C}^{N}$ denote the local beamforming direction at AP $l$ for subgroup $g$. The transmit precoder is defined as
\begin{equation}
    \mathbf{w}_{lg} =
    \sqrt{\rho_{lg}} \,
    \frac{\mathbf{v}_{lg}}
    {\sqrt{\mathbb{E}\{\|\mathbf{v}_{lg}\|^{2}\}}},
    \label{eq:precoder_norm}
\end{equation}
where $\rho_{lg}$ is the DL power allocated by AP $l$ to subgroup $g$,
and the normalization ensures
$\mathbb{E}\{\|\mathbf{w}_{lg}\|^{2}\}=\rho_{lg}$.

\subsection{Classical Conjugate Beamforming (CB)}

Classical CB uses the MMSE composite estimate as the beamforming direction as
\begin{equation}
    \mathbf{v}_{lg}^{\mathrm{CB}} = \hat{\mathbf{h}}_{l}^g.
\end{equation}
Thus, the CB precoder at AP $l$ is
\begin{equation}
    \mathbf{w}_{lg}^{\mathrm{CB}} =
    \sqrt{\rho_{lg}}\,
    \frac{\hat{\mathbf{h}}_{l}^g}
    {\sqrt{\mathbb{E}\{\|\hat{\mathbf{h}}_{l}^g\|^{2}\}}},
    \label{eq:CB}
\end{equation}
with
$\mathbb{E}\{\|\hat{\mathbf{h}}_{l}^g\|^{2}\}
= \mathrm{tr}\!\left(K_g^2 \mathbf{R}^{g}_{l}
\boldsymbol{\Gamma}_g^{-1}\mathbf{R}^{g}_{l}\right)$.
CB has the lowest computational complexity and maximizes the coherent array gain, but exhibits limited channel hardening due to the heterogeneity of AP‑user distances in CF‑mMIMO deployments.

\subsection{Normalized Conjugate Beamforming (NCB)}

NCB improves hardening by normalizing the beamforming direction by its instantaneous norm \cite{2020Femenias}:
\begin{equation}
    \mathbf{v}_{lg}^{\mathrm{NCB}}
    = \frac{\hat{\mathbf{h}}_{l}^g}
           {\|\hat{\mathbf{h}}_{l}^g\|}.
\end{equation}
Since $\mathbb{E}\{\|\mathbf{v}_{lg}^{\mathrm{NCB}}\|^{2}\}=1$, the DL precoder simplifies to
\begin{equation}
    \mathbf{w}_{lg}^{\mathrm{NCB}}
    = \sqrt{\rho_{lg}}
      \frac{\hat{\mathbf{h}}_{l}^g}
           {\|\hat{\mathbf{h}}_{l}^g\|}.
    \label{eq:NCB}
\end{equation}
NCB enforces instantaneous per‑AP power constraints and significantly reduces the variance of the effective DL channel, providing stronger channel hardening than CB, particularly when $N$ is moderate or large.

\subsection{Enhanced Conjugate Beamforming (ECB)}

ECB further enhances the channel hardening effect by normalizing by the \emph{squared} instantaneous norm \cite{2021Interdonato}:
\begin{equation}
    \mathbf{v}_{lg}^{\mathrm{ECB}}
    = \frac{\hat{\mathbf{h}}_{l}^g}
           {\|\hat{\mathbf{h}}_{l}^g\|^{2}}.
\end{equation}
The resulting precoder is
\begin{equation}
    \mathbf{w}_{lg}^{\mathrm{ECB}} =
    \sqrt{\rho_{lg}}\,
    \frac{\hat{\mathbf{h}}_{l}^g}
         {\|\hat{\mathbf{h}}_{l}^g\|^{2}}
    \frac{1}
    {\sqrt{\mathbb{E}\{\|\hat{\mathbf{h}}_{l}^g\|^{-2}\}}},
    \label{eq:ECB}
\end{equation}
where the factor
$\frac{1}{\sqrt{\mathbb{E}\{\|\hat{\mathbf{h}}_{l}^g\|^{-2}\}}}$ is computed offline
via Monte Carlo averaging. ECB offers the strongest hardening among the
three schemes and is particularly advantageous when the AP antenna array
is sufficiently large.

\subsection{Distributed Power Allocation}

To satisfy the per‑AP power constraint, we adopt an adaptive power
allocation (APA) rule consistent with scalable CF‑mMIMO design:
\begin{equation}
    \rho_{lg} =
    \begin{cases}
        \displaystyle
        P_{\mathrm{dl}}
        \frac{
        \left( \mathrm{tr}\,\mathbf{R}^{g}_{l} \right)^{\nu}
        }{
        \sum_{c\in\mathcal{D}_{l}}
        \left( \mathrm{tr}\,\mathbf{R}^{c}_{l} \right)^{\nu}
        },
        & g\in \mathcal{D}_{l}, \\[3mm]
        0, & \text{otherwise},
    \end{cases}
    \label{eq:apa}
\end{equation}
where $\mathcal{D}_{l}$ is the set of subgroups served by AP $l$ and
$\nu$ tunes the fairness/performance trade‑off.



\section{Numerical Results}
\label{sec:results}
\subsection{Simulation Setup}

This section presents numerical results to evaluate the performance of subgroup-centric CF-mMIMO multicasting using distributed CB variants. Monte Carlo simulations are conducted to assess classical CB, NCB, and ECB in terms of aggregated SE (ASE).

A CF-mMIMO network composed of $L=100$ distributed APs, each equipped with $N=\{4,8,16\}$ antennas, serving $K=100$ single-antenna UEs over the same time-frequency resources is considered. APs and UEs are deployed over a square area of side $1000$ m with wrap-around to avoid boundary effects. 

The coherence block length is set to $\tau_{\mathrm{c}}=200$ samples, with $\tau_{\mathrm{p}}=\min(G,20)$ UL pilot symbols and $\tau_{\mathrm{d}}=\tau_{\mathrm{c}}-\tau_{\mathrm{p}}$ DL data symbols. All UEs transmit UL pilots with equal power, $P_{\mathrm{p}}=100$~mW, and the total DL transmit power per AP is constrained and equally applied across all schemes, $P_{\mathrm{dl}}=200$~mW. 
Each reported \gls{cdf} corresponds to an average over $500$ independent random network realizations and $100$ small-scale fading channel realizations per deployment.

\subsection{Uniform User Distribution}

We first consider a scenario where the $K=100$ UEs are uniformly distributed across the coverage area, which is modeled by deploying $100$ clusters with a single UE per cluster. This geometry 
makes UEs experience significantly different large-scale fading conditions with respect to the distributed APs.

Fig.~2 shows the CDF of the ASE achieved by CB, NCB, and ECB for $G=1$ (pure multicast), $G=10$ (subgroup multicast), and $G=100$ (unicast), and for different numbers of antennas per AP. In this scenario, unicast transmission ($G=100$) clearly dominates across the entire CDF, achieving the highest ASE for all precoding schemes and antenna configurations. This behavior confirms that, under uniformly distributed user deployments, the system can fully exploit spatial multiplexing gains.

Among the CB-based precoders, NCB systematically improves upon classical CB by reducing the variability of the effective channel gain through instantaneous normalization. ECB provides additional gains only when the number of antennas per AP is sufficiently large, whereas for small and moderate antenna arrays, the advantages of squared-norm normalization are limited. Subgroup multicast ($G=10$) improves upon pure multicast ($G=1$), but remains significantly outperformed by unicast transmission in this geometry.

\subsection{Spatially Clustered User Deployments}

\subsubsection{Moderately Clustered Users}

We next analyze a moderately clustered deployment in which the $100$ UEs are grouped into $10$ spatial clusters\footnote{Note that spatial clusters deployed in this paper correspond to square areas (hotspots) of side $10$ m.} of $10$ UEs each. In this setting, UEs within the same cluster experience very similar propagation conditions, while different clusters remain spatially separated.

Fig.~3 reports the ASE CDFs for this scenario. Compared to the uniform case, unicast transmission suffers from a pronounced performance degradation due to increased pilot contamination and strong intra-cluster interference. In contrast, subgroup-based multicasting with $G=10$ achieves the highest ASE across most realizations,
highlighting the effectiveness of subgroup-centric transmission in moderately clustered environments.

Moreover, classical CB and NCB exhibit almost identical performance, indicating that the benefits of instantaneous normalization are greatly reduced in clustered geometries. ECB does not provide noticeable ASE gains for any of the considered antenna configurations, which suggests that the channel hardening conditions required to exploit its normalization are not sufficiently met in this scenario.

\subsubsection{Highly Concentrated User Deployment}

Finally, we consider an extreme clustering scenario in which all $100$ UEs are concentrated within a single spatial cluster. This deployment models dense crowd scenarios where UEs experience nearly identical large-scale fading conditions.

The corresponding ASE distributions are shown in Fig.~4. In this case, unicast transmission collapses almost completely due to severe pilot contamination, yielding negligible ASE values. Multicast transmission becomes essential, with subgroup multicast ($G=10$) and pure multicast ($G=1$) significantly outperforming unicast.

In this regime, classical CB emerges as the most robust precoding strategy, slightly outperforming NCB and ECB across the entire CDF. While increasing the number of antennas per AP improves the absolute ASE, it does not change the relative ranking of the precoders. These results indicate that, under extreme clustering, simple and fully distributed CB provides an effective and scalable solution.

\begin{figure}[t]
    \centering
    \includegraphics[width=\columnwidth]{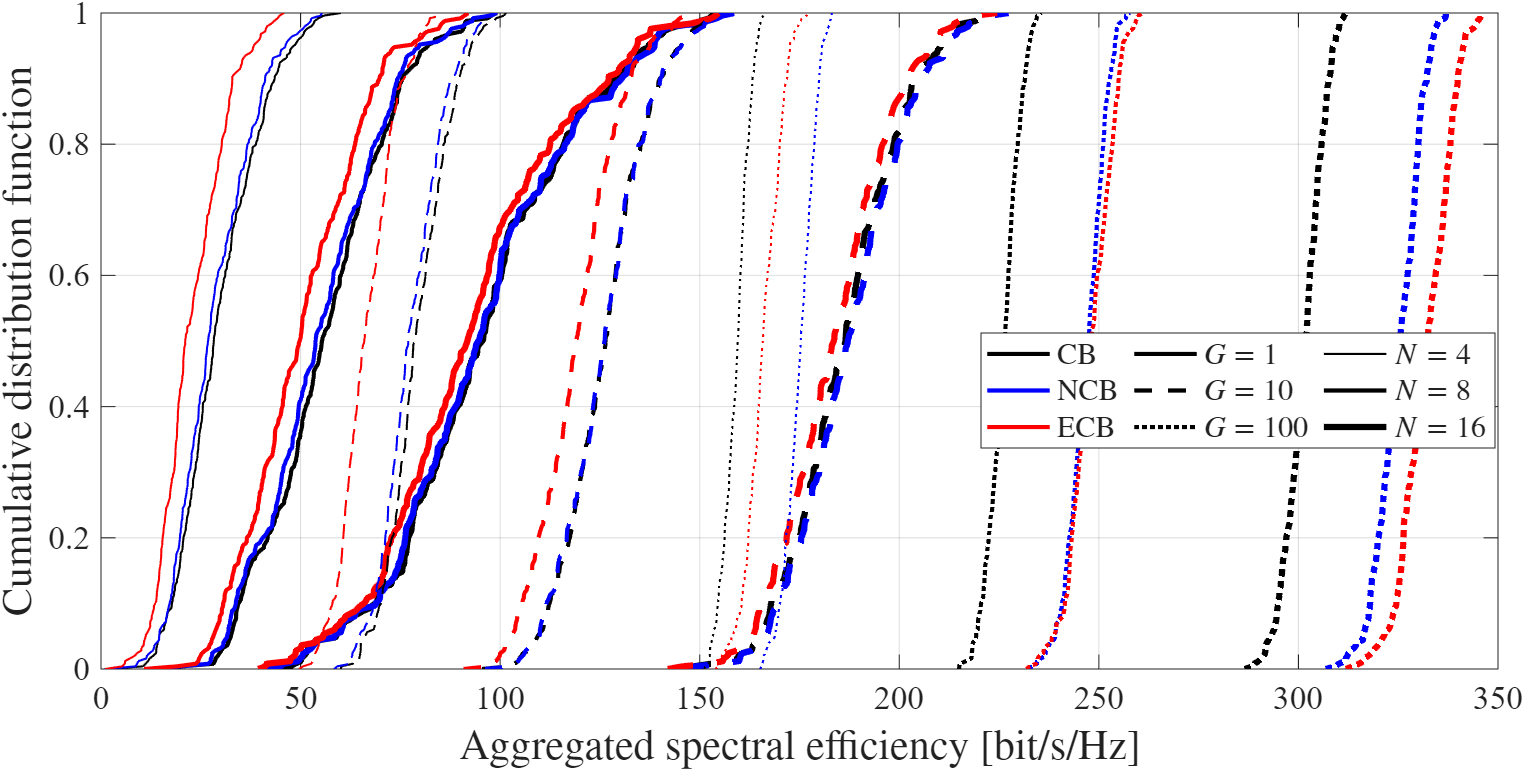}
    \caption{CDF of the aggregated spectral efficiency for a uniform user distribution (100 clusters of 1 user), comparing CB, NCB, and ECB precoders for $G=\{1,10,100\}$ and $N=\{4,8,16\}$ antennas per AP. Colors indicate the precoding scheme, line styles denote the number of subgroups, and line thickness represents the number of antennas
    per AP.}
    \label{fig:uniform}
\end{figure}

\begin{figure}[t]
    \centering
    \includegraphics[width=\columnwidth]{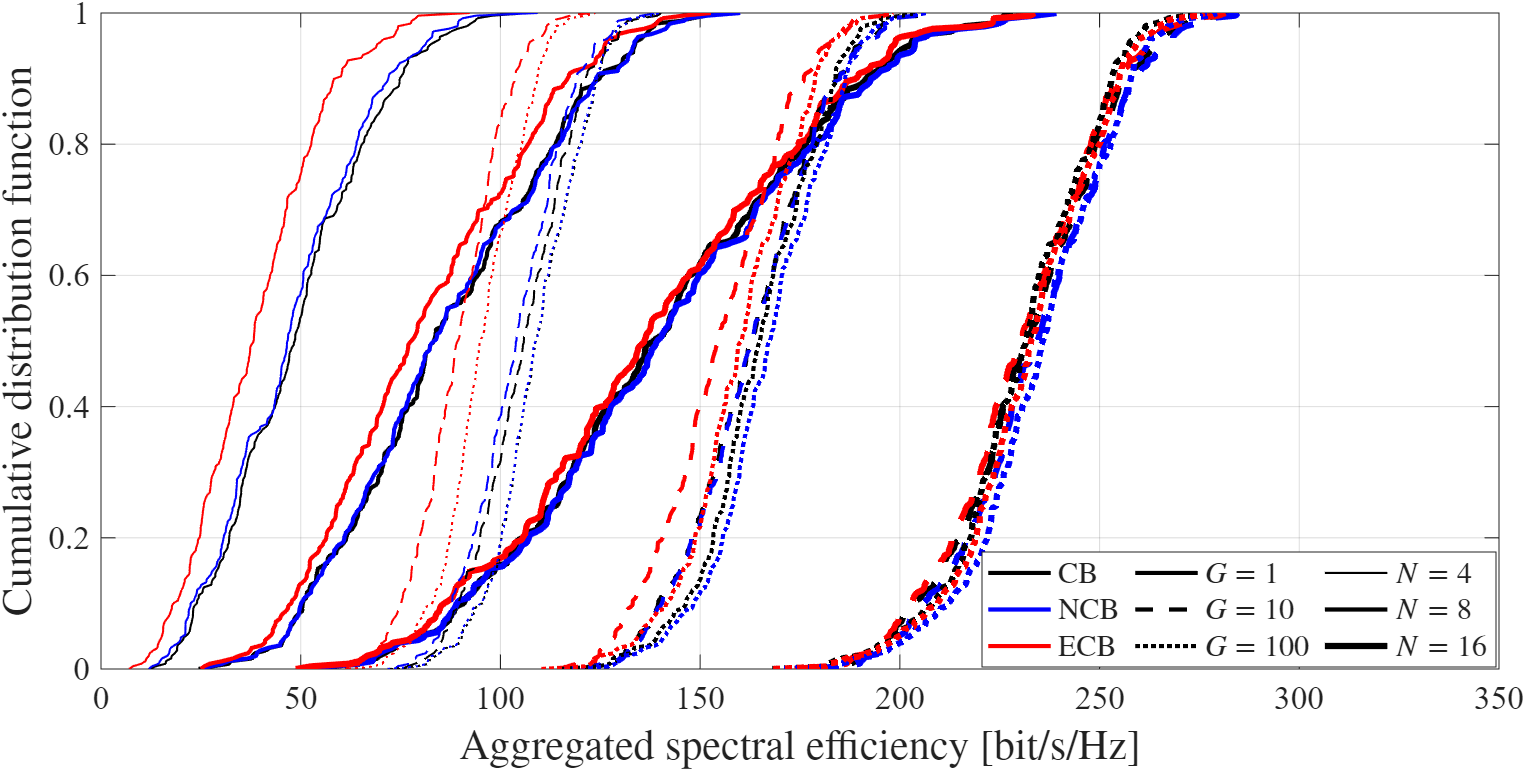}
    \caption{CDF of the aggregated spectral efficiency for a moderately
    clustered deployment (10 clusters of 10 users), comparing CB, NCB,
    and ECB for $G=\{1,10,100\}$ and $N=\{4,8,16\}$. Subgroup multicast
    ($G=10$) emerges as the preferred transmission mode.}
    \label{fig:moderate}
\end{figure}

\begin{figure}[t]
    \centering
    \includegraphics[width=\columnwidth]{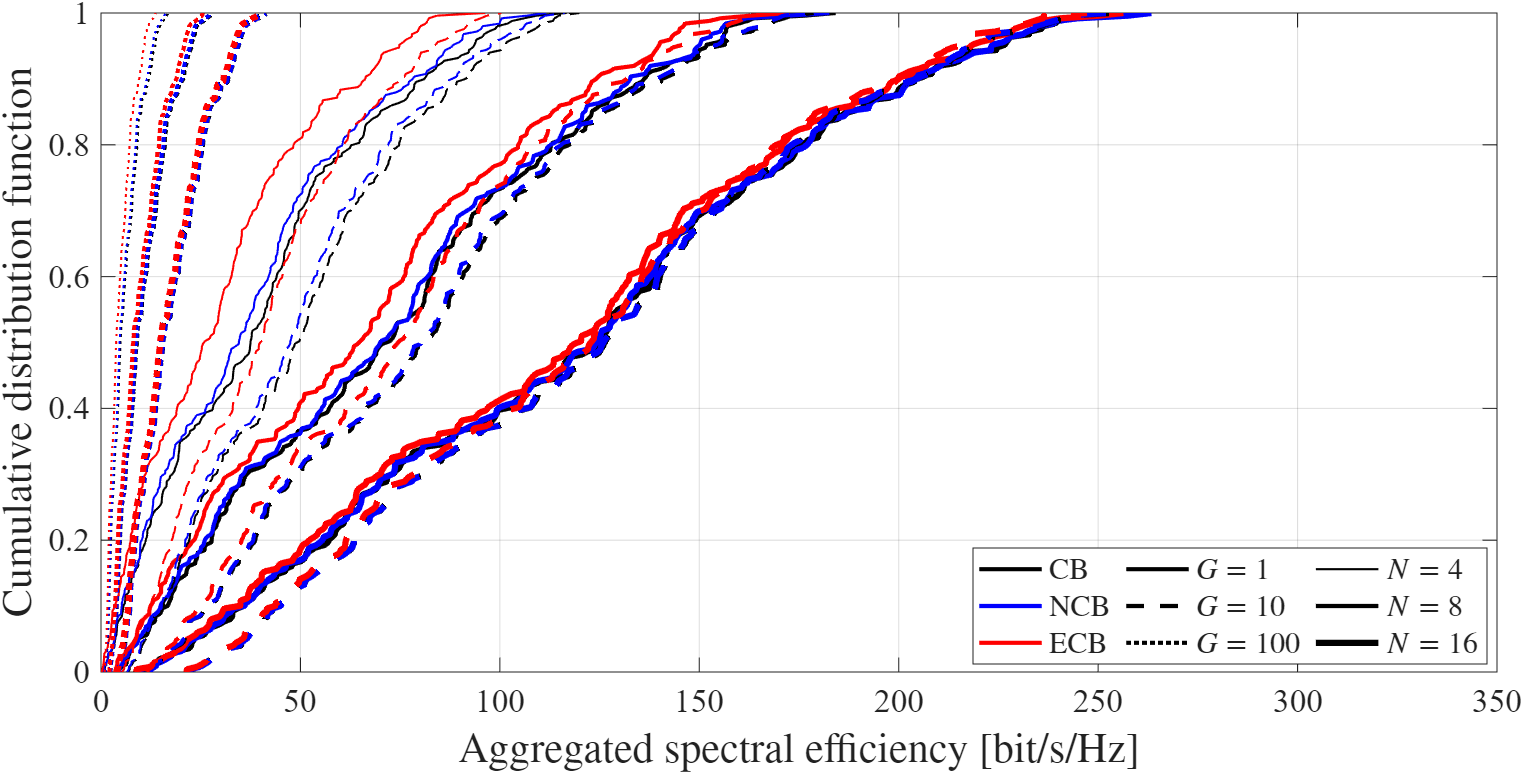}
    \caption{CDF of the aggregated spectral efficiency for a highly
    concentrated user deployment (single cluster of 100 users).
    Unicast transmission collapses, while multicast-based schemes remain
    effective.}
    \label{fig:extreme}
\end{figure}

\subsection{Heterogeneous User Deployment Scenarios}

In practical deployments, multicast users are rarely either fully uniformly distributed or completely concentrated in spatial clusters. Instead, realistic scenarios typically combine isolated users with dense
crowds or hotspots. To capture such conditions, we investigate \emph{heterogeneous user deployments} that mix uniformly distributed users with spatial clusters of different sizes.
Unlike the previous uniform and clustered scenarios with $K=100$, we adopt the heterogeneous deployment model of~\cite{2024delaFuente} and consider $K=500$ UEs distributed over a square area of side $1000$~m.
Across all scenarios, the total number of UEs remains constant; only the \emph{composition} between uniformly distributed UEs and clustered UEs
varies. This modeling choice enables a systematic assessment of how spatial heterogeneity impacts both the achievable ASE and the optimal number of
multicast subgroups.

\textit{Scenario~1 (highly clustered):}
The deployment is dominated by dense UE clusters, with only a small
uniform component.
Specifically, $10$ UEs are uniformly distributed over the coverage area,
while the remaining UEs form spatial clusters composed of two clusters of
$10$ UEs, four clusters of $20$ UEs, three clusters of $30$ UEs, and six
clusters of $50$ UEs.
This scenario models strongly hotspot-driven environments such as
stadiums or large public events.

\textit{Scenario~2 (moderately clustered):}
A more balanced heterogeneous deployment is considered.
In this case, $20$ UEs are uniformly distributed, while the remaining UEs
are grouped into two clusters of $10$ UEs, three clusters of $20$ UEs,
five clusters of $30$ UEs, and five clusters of $50$ UEs.
This scenario represents mixed environments where isolated UEs coexist
with several medium-sized hotspots.

\textit{Scenario~3 (sparsely clustered):}
Most UEs are uniformly distributed across the coverage area, and only a
limited number of spatial clusters are present.
Specifically, $100$ UEs are uniformly distributed, while the remaining
UEs form five clusters of $10$ UEs, five clusters of $20$ UEs, five
clusters of $30$ UEs, and two clusters of $50$ UEs.
This scenario approaches a near-uniform deployment while still retaining localized concentrations of UEs.

\begin{figure}[!t]
\subfloat[CB precoding]
{\includegraphics[width=\columnwidth]
{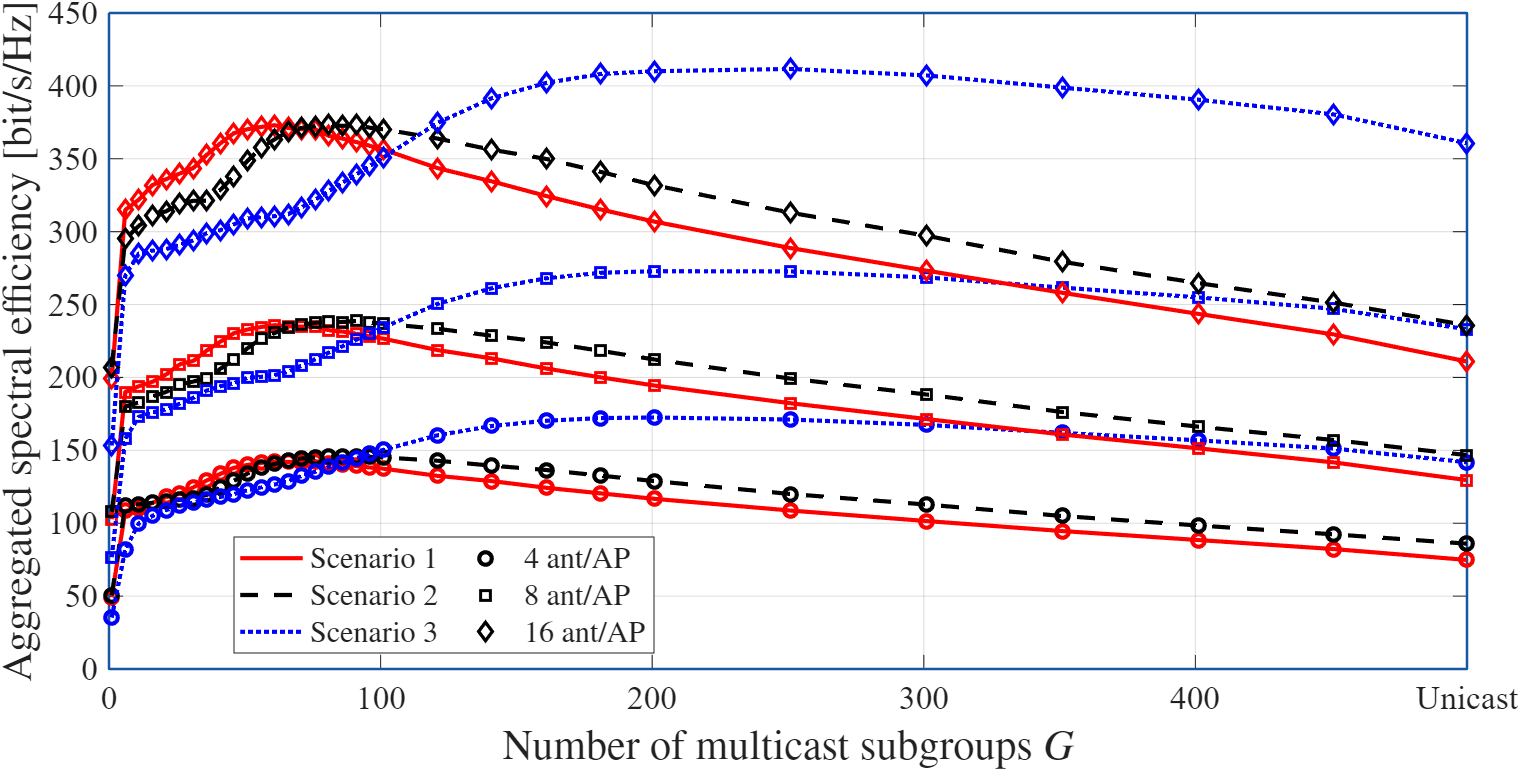}\label{fig:CB-het}}\\
\subfloat[NCB precoding]
{\includegraphics[width=\columnwidth]
{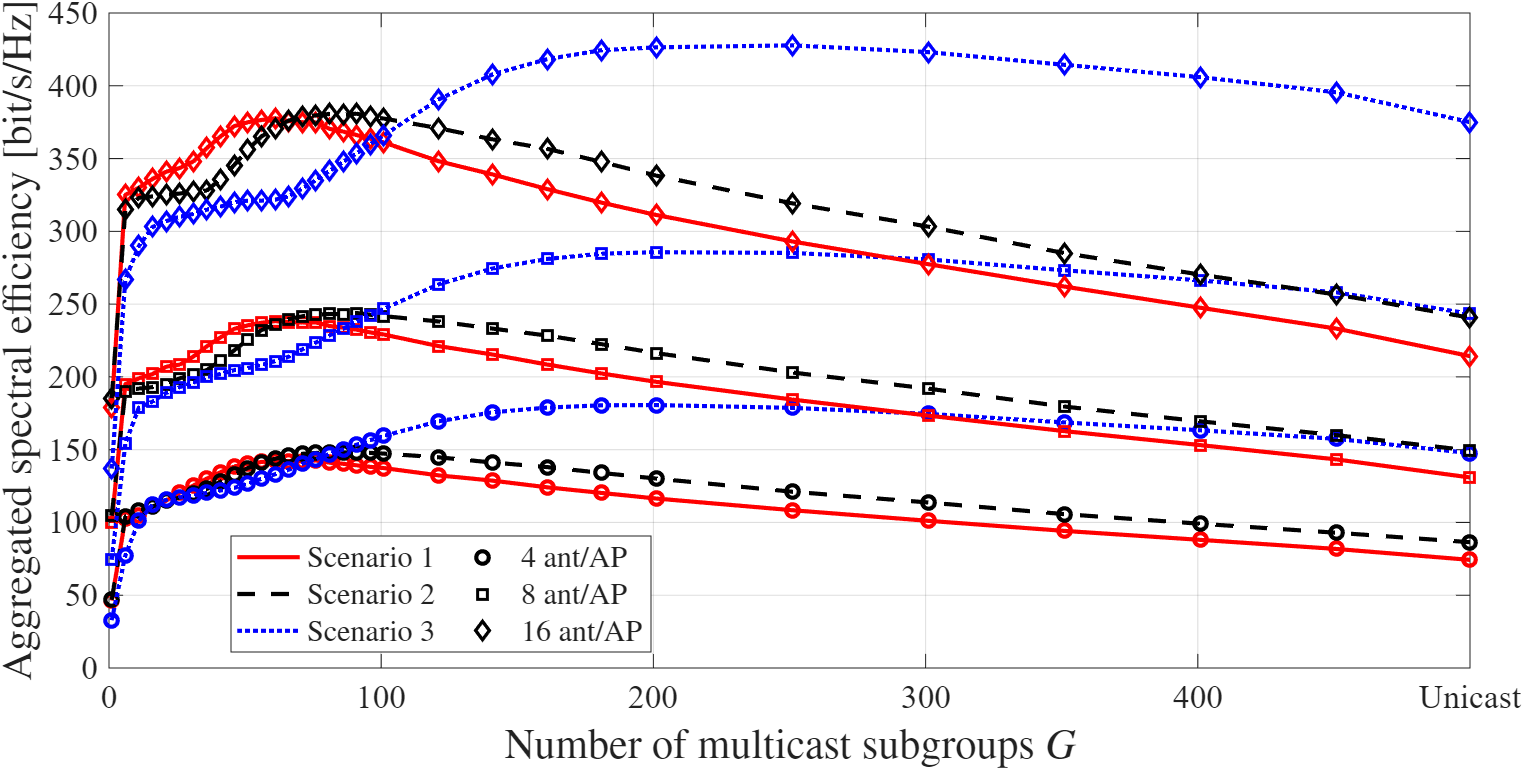}\label{fig:NCB-het}}\\
\subfloat[ECB precoding]
{\includegraphics[width=\columnwidth]
{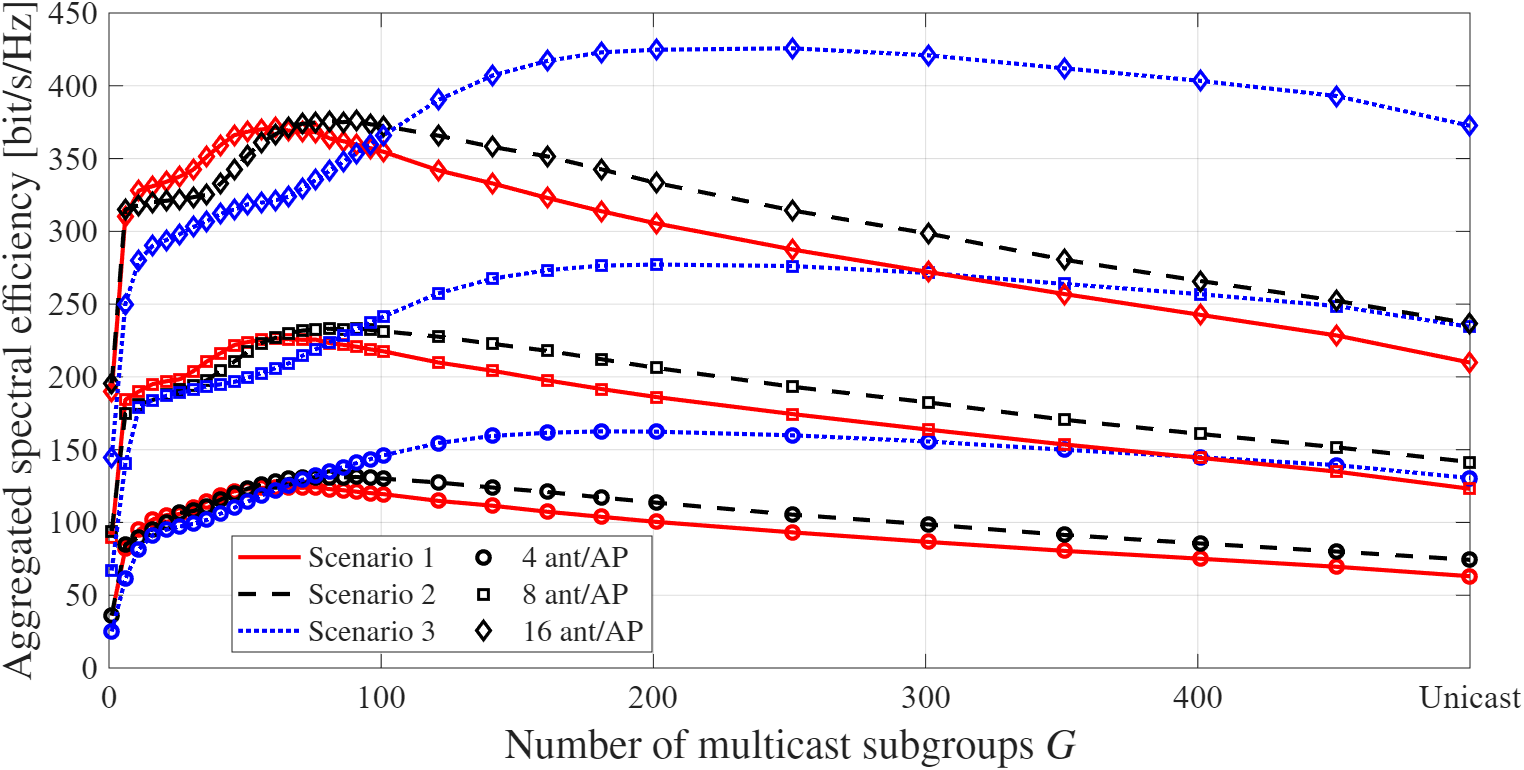}\label{fig:ECB-het}}
\caption{Aggregated spectral efficiency versus the number of multicast
subgroups $G$ for heterogeneous user deployments. Three scenarios are
considered, ranging from highly clustered (Scenario~1) to sparsely
clustered (Scenario~3). Results are shown for distributed CB, NCB,
and ECB precoding, and for $N=\{4,8,16\}$ antennas per AP.}
\label{fig:heterogeneous}
\end{figure}

These three scenarios allow us to analyze how the optimal subgrouping strategy evolves as the UE distribution transitions from cluster-dominated to almost uniform conditions.
Fig.~\ref{fig:heterogeneous} reports the ASE as a function of the number of multicast subgroups $G$ for the three heterogeneous scenarios, considering
distributed CB, NCB, and ECB precoding.
Each subfigure corresponds to a different precoding scheme, while curves with distinct markers indicate the number of antennas per AP.

Fig.~\ref{fig:CB-het} shows the results obtained with distributed CB
precoding.
For all heterogeneous deployments, the ASE exhibits a clear unimodal
behavior with respect to $G$, with a well-defined maximum at intermediate
subgroup counts.
This confirms that neither single-group multicast nor pure unicast
transmission achieves optimal performance in heterogeneous environments.
The ASE-maximizing value of $G$ strongly depends on the degree of UE
clustering: in highly clustered scenarios, the optimum is attained with a
relatively small number of multicast subgroups, whereas in moderately and
sparsely clustered deployments the optimal $G$ progressively shifts
toward larger values as the fraction of uniformly distributed UEs
increases.

A similar dependence of the ASE on $G$ is observed for both NCB and ECB
precoding, as shown in Figs.~\ref{fig:NCB-het}
and~\ref{fig:ECB-het}.
In all heterogeneous scenarios, the ASE curves remain unimodal with
respect to $G$, and the location of the ASE-maximizing subgroup count is
mainly dictated by the spatial distribution of the UEs.
However, Fig.~5 also indicates that, unlike the subgrouping behavior, the
absolute ASE levels achieved by CB, NCB, and ECB depend on the specific
operating regime.
In particular, NCB and ECB do not systematically outperform CB across all
scenarios and antenna configurations.
While ECB can provide noticeable gains when a large number of antennas
per AP is available (e.g., $N=16$) and the UE distribution is weakly
clustered, such gains diminish or vanish for smaller antenna arrays and
more clustered deployments.
Overall, these results show that the choice of precoding scheme influences the achievable ASE, but does not alter the location of the optimal
subgrouping point.

The heterogeneous-deployment results demonstrate that multicast subgrouping constitutes a robust and adaptable transmission strategy
capable of effectively bridging clustered and nearly uniform UE distributions.
The optimal number of multicast subgroups is primarily governed by the spatial geometry of the UEs, whereas the precoding scheme mainly impacts
the absolute ASE levels in a scenario- and antenna-dependent manner.
This observation highlights the relevance of subgroup-centric multicast transmission for practical CF-mMIMO deployments.

\section{Conclusion}

This paper has investigated scalable CB variants for physical-layer multicasting in CF-mMIMO systems under subgroup-centric transmission.
By comparing classical CB, NCB, and ECB in fully distributed settings, the results show that multicast performance is primarily driven by the
spatial geometry of the users.

Uniform UE deployments favor unicast transmission, where NCB and ECB (when sufficiently large antenna arrays are available) can offer gains over classical CB. In contrast, for spatially clustered deployments,
unicast transmission becomes inefficient due to pilot contamination, and subgroup-based multicasting emerges as the preferred operating mode, with classical CB providing robust and competitive performance.

Heterogeneous deployments further confirm that no single precoding scheme is universally optimal. While the choice of precoder mainly affects the achievable ASE level, the optimal number of multicast
subgroups is largely dictated by the degree of spatial heterogeneity. These results highlight the importance of jointly selecting the transmission mode, subgrouping level, and precoding strategy based on large-scale spatial characteristics.

As future work, an interesting direction is the extension to heterogeneous CF-mMIMO systems with APs equipped with different numbers of antennas. In such scenarios, adaptive and distributed schemes that allow each AP to select, on a per-subgroup basis, between CB, NCB, and ECB based on local parameters such as the number of UEs in the subgroup or the number of antennas in the AP could further improve performance while preserving scalability.

\bibliographystyle{IEEEtran}
\bibliography{IEEEabrv,main}
\end{document}